\documentclass[12pt]{article}

\newcommand{\beq}{\begin{equation}}
\newcommand{\eeq}{\end{equation}}
\newcommand{\beqa}{\begin{eqnarray}}
\newcommand{\eeqa}{\end{eqnarray}}
\newcommand{\beqar}{\begin{eqnarray*}}
\newcommand{\eeqar}{\end{eqnarray*}}

\newcommand{\al}{\alpha}
\newcommand{\be}{\beta}

\def\Tr           {\mbox{\rm Tr}\,}

\def\cd           {{\cdot}}

\def\fsC    {C\!\!\!\!/\,}
\def\fsH    {H\!\!\!\!/\,}

\newcommand{\eps}{\epsilon}

\newcommand{\lam}{\lambda}

\newcommand{\z}{\zeta}

\newcommand{\labell}[1]{\label{#1}} 
\newcommand{\reef}[1]{(\ref{#1})}
\newcommand\prt{\partial}
\newcommand\veps{\varepsilon}

\def\sst#1{{\scriptscriptstyle #1}}
\def\0{{\sst{(0)}}}
\def\1{{\sst{(1)}}}
\def\2{{\sst{(2)}}}
\def\3{{\sst{(3)}}}
\def\4{{\sst{(4)}}}
\def\5{{\sst{(5)}}}
\def\6{{\sst{(6)}}}
\def\7{{\sst{(7)}}}
\def\8{{\sst{(8)}}}

\begin{document}
\baselineskip 18pt%
\begin{titlepage}
\vspace*{1mm}%
 
 TUW-16-19

\vspace*{8mm}
\vspace*{8mm}%

\center{ {\bf \Large Remarks on Non-BPS String Amplitudes and Their All Order $\alpha'$ Contact Interactions in IIB, IIA}} 

\begin{center}
{Ehsan Hatefi   }\footnote{
ehsan.hatefi@tuwien.ac.at, e.hatefi@qmul.ac.uk, ehsan.hatefi@cern.ch, ehsanhatefi@gmail.com }

\vspace*{0.6cm}{ Institute for Theoretical Physics, TU Wien
\\
Wiedner Hauptstrasse 8-10/136, A-1040 Vienna, Austria}
\vskip.06in

\vspace*{.3cm}
\end{center}
\begin{center}{\bf Abstract}\end{center}
\begin{quote}

We explore the entire form of S-Matrix elements of a potential  $C_{n-1}$ Ramond-Ramond  (RR) form field, a tachyon and two transverse scalar fields  on both world volume and transverse directions  of type IIB and IIA superstring theories. Apart from $<V_{C^{-2}}V_{\phi^{0}}V_{\phi ^{0}}V_{T ^{0}}>$ the other scattering amplitude, namely
$<V_{C^{-1}}V_{\phi^{-1}}V_{\phi ^{0}}V_{T ^{0}}>$ is also revealed. We then start to compare all singularity structures of symmetric and asymmetric analysis, generating all infinite singularity structures as well as all order $\alpha'$ contact interactions on the whole directions. This leads to deriving various new contact terms and several new restricted Bianchi identities in both type IIB and IIA. It is also shown that just some of the new couplings  of type IIB (IIA) string theory can be re-verified in an Effective Field Theory (EFT) by pull-back of branes. To construct the rest of S-matrix elements one needs to first derive restricted world volume (or bulk) Bianchi identities and then discover new  EFT couplings in both  type IIB and IIA. Finally the presence of commutator of scalar fields inside the exponential of Wess-Zumino action for non-BPS branes has been confirmed as well.

 \end{quote}
\end{titlepage}

 \section{Introduction}

Over the last two decades D-branes or the so called fundamental objects have been at the center of attention.  Without no doubt  the presence of theses objects has opened up  enormous potentials in various subjects in theoretical high energy physics,  most specifically in type IIA,IIB superstring theories \cite{Polchinski:1995mt,Witten:1995im}. 
\vskip.1in

In order to address the dynamics of D-branes, one has to come up with clear understanding of how to properly reformulate or restructure the effective actions of D-branes.  The mix combinations of closed string Ramond-Ramond and two transverse scalar field couplings including the appearance of their commutator have been investigated by the so called dielectric effect. Indeed the simplest  effective actions for multiple branes have been long proposed by Myers \cite{Myers:1999ps}. The next step was to achieve the generalization of Myers action for different field content as well as embedding all  order $\alpha'$ corrections within a compact formula. Although they are complicated,  these tasks have also been answered by dealing with direct conformal field techniques as well as higher point functions of  (non)-BPS amplitudes. In fact the generalization of Myers action for non-Abelian cases with its all order $\alpha'$ corrections have been explored in \cite{Hatefi:2012zh}.

\vskip.06in

As it quite often happens in string theory, among other couplings one expects to predict some anomalous couplings. One might wonder  how branes could potentially dissolve through themselves  where  these remarks up to  some satisfaction \cite{Green:1996dd} have also been answered in detail. It has been highlighted  that the three standard formalisms  of obtaining Effective Field Theory (EFT) couplings such as Myers terms, pull back of D-branes and Taylor expansion are not good enough to produce string amplitudes. Indeed recently in \cite{Hatefi:2015ora} we have shown that not only scattering amplitudes will have strong potential to gain new couplings with  new structures but also they are able to exactly fix  the coefficients of those couplings without any ambiguity.  One can go further to find out  all order $\alpha'$ corrections  to those new discovered couplings as well \cite{Hatefi:2016wof}.

\vskip.1in

One might want to look for some real physical applications to those new actions that have been explored so far. For instance, we would like to hint on $N^3$ phenomenon of $M5$ branes as well as different  configurations, like M2-M5 systems . Another example that can be raised,  has to do with dS solutions or understanding how entropy of different systems is  scaled by directly employing the commutator of scalar fields \cite{Hatefi:2012bp}. Let us promptly go through some literatures of effective actions.

\vskip.1in

Non-Abelian symmetric trace action was imposed in \cite{Howe:2006rv} and long before that the action for  single bosonic brane had been proposed by R.Leigh  in \cite{Leigh:1989jq}.  Later on its supersymmetric partner effective action  was derived by \cite{Cederwall:1996pv}.  On the other hand, various people including the author started addressing  the entire form of  the effective actions. More importantly, we tried to figure out what needs to be pursued  to obtain not only non-BPS brane's effective actions but also D-brane anti D-brane  effective actions  of string theory, involving their all order $\alpha'$ corrections. It is worth to point out that these actions have to come out of consistent comparisons of EFT couplings with string computations.  Due to the presence of tachyons,  we no longer can even rely on the notion of duality \cite{Michel:2014lva}.
\vskip.1in

We end the vast literatures by just mentioning  a worthwhile comment.  With scattering amplitude prospective, RR  couplings with non-BPS branes have also been investigated \cite{Kennedy:1999nn}. Among other things,  one of our goals of this paper is to actually find out RR couplings in asymmetric picture.  We intend not to write down the actions that have been explored, and would rather read off  some of the  EFT methods and their effective actions from \cite{Hatefi:2012wj}.

\vskip.1in

The outline of the paper is  such that, we derive  the entire form of a potential C-field, two transverse scalar fields and a tachyon  $<V_{C^{-2}}  V_{\phi^{0}} V_{\phi ^{0}}  V_{T ^{0}}>$ as well as  $<V_{C^{-1}}  V_{\phi^{-1}} V_{\phi ^{0}}  V_{T ^{0}}>$ on both IIA,IIB  superstring theories. 
Although recent points have been raised  by \cite{Sen:2015hia}, obviously we want to address the complete S-matrix elements on both bulk and world volume directions. More specifically, we would like to clearly consider the terms in our scattering amplitude  that involve the scalar products of RR momenta and the polarizations of transverse scalar fields, that is,  
$p.\xi_1$ and  $p.\xi_2$ terms. 

\vskip.1in

 We obtain some restricted Bianchi identities on both bulk and world volume directions. We also show that unlike  $<V_{C^{-2}}  V_{\phi^{0}} V_{\phi ^{0}}  V_{A ^{0}}>$ for this particular non-BPS amplitude,  there are no u,s channel bulk singularity structures. All infinite t, $(t+s'+u')$  singularities have been shown to match with an EFT counterpart. By dealing with direct string amplitude, we also confirm the presence of  new EFT couplings in both IIA,IIB \footnote{ Since the traces of string amplitudes are non zero for different $p,n$ cases, all these new EFT couplings are valid for both IIA,IIB superstring theories.} as follows
 \beqa
 &&(2\pi\alpha')^3\mu'_p \beta' \int_{\Sigma_{p+1}} d^{p+1} \sigma  \Tr(\partial_iC^{j}_{p-1} \wedge D\phi_j\wedge D T\phi^i),  \nonumber\\&&  (2\pi\alpha')^3\mu'_p \beta' \int_{\Sigma_{p+1}} d^{p+1} \sigma\Tr(C_{p}\wedge DT \phi^i\phi_i), \nonumber\\&&
(2\pi\alpha')^3\mu'_p \beta'\int_{\Sigma_{p+1}} d^{p+1} \sigma\Tr(C_{p-2} \wedge D\phi^i\wedge D\phi_i\wedge DT), \nonumber\\&& (2\pi\alpha')^3\mu'_p \beta' \int_{\Sigma_{p+1}} d^{p+1} \sigma
  \Tr(D_{a_0}T[\phi^i,\phi^j])C_{jia_1\cdots a_p}\veps^{a_0\cdots a_p}.  
 \eeqa
 where none of these couplings comes from standard effective field theory methods nor from pull-back, Taylor expansion. Notice to the important point as follows. 
 \vskip.1in
 
 We also clarify  the structures of new couplings on the entire space-time dimensions, fixing their coefficients. More crucially,  the generalization of their all order $\alpha'$ higher derivative corrections without any ambiguity will be constructed as well. The presence of the commutator of  two transverse scalar fields from the exponential of Wess-Zumino action even for non-BPS branes will be proven later on. The notation is such that $\mu,\nu = 0, 1,..., 9$, world volume and transverse directions are accordingly  covered by $ a,b, c = 0, 1,..., p$ and $i,j = p + 1,...,9$.
 
 \section{Entire non-BPS $<C^{-2} \phi^0  \phi^0 T^0>$ scattering amplitude}
 
To investigate the whole scattering amplitude of a $(p+1)$- potential form field $C$, two massless scalar fields and a tachyon $<C^{-2} \phi^0  \phi^0 T^0>$ as well as all singularity and contact term structures, we need to work out all Conformal Field Theory (CFT) methods.

The structures of vertex operators including their Chan-Paton factors for non-BPS branes are given in \cite{Hatefi:2013yxa} as follows
\beqa
V_{T}^{(0)}(x) &=& i\alpha'k.\psi(x) e^{2ik\cd X(x)}\lam\otimes\sigma_1,
\nonumber\\
V_{\phi}^{(0)}(x) &=& \xi_{i}\bigg(\partial X^i(x)+i\alpha'q\cd\psi\psi^i(x)\bigg)e^{2iq.X(x)}\lam\otimes I,
\nonumber\\
V_{\phi}^{(-1)}(y) &=& \xi_{i}e^{-\phi(y)}\psi^i(y)e^{2iq.X(y)}\lam\otimes \sigma_3,
\nonumber\\
V_{RR}^{(-1)}(z,\bar{z})&=&(P_{-}\fsH_{(n)}M_p)^{\al\be}e^{-\phi(z)/2} S_{\al}(z)e^{ip\cd X(z)}e^{-\phi(\bar{z})/2} S_{\be}(\bar{z}) e^{ip\cd D \cd X(\bar{z})}\otimes\sigma_3\sigma_1,\nonumber\\
V_{RR}^{(-2)}(z,\bar{z})&=&(P_{-}\fsC_{(n-1)}M_p)^{\al\be}e^{-3\phi(z)/2} S_{\al}(z)e^{ip\cd X(z)}e^{-\phi(\bar{z})/2} S_{\be}(\bar{z}) e^{ip\cd D \cd X(\bar{z})}\otimes\sigma_1.
\label{vertex}\eeqa

 For  brane-anti brane system we have a non-zero coupling between an RR and two tachyons as  $<C^{-1} T^{-1} T^0>$ makes sense. It is shown in  \cite{Hatefi:2013yxa} that tachyon in zero and (-1) picture carries $\sigma_1$ and $\sigma_2$ CP factor accordingly. Hence, RR  in $(-1/2, -1/2)$ picture for brane-anti brane system must carry   $\sigma_3$ CP matrix. We have also discussed in \cite{Hatefi:2012wj} that for non-BPS branes  in (-1) picture there has to be another extra coefficient  so that RR in (-1) picture for the non-BPS branes does carry  $\sigma_3\sigma_1$ CP matrix. Picture changing operator (PCO) also carries $\sigma_3$ CP factor. If we apply PCO to RR in (-1) picture , we then get to know that RR in (-2) picture carries $\sigma_1$ CP factor.

 The vertex of RR  was releazed in \cite{Bianchi:1991eu} and \cite{Liu:2001qa} appropriately.  One might find various results of BPS amplitudes in \cite{Barreiro:2013dpa}. Given the literatures, we just start to look after all the correlation functions.\footnote{ Note that since winding modes can not be embedded for entire 10 dimensions flat space, we should carry out direct $<V_{C^{-2}} V_{\phi^0} V_{\phi^0}V_{T^0}>$ analysis and there is no point to compare it with $<V_{C^{-2}} V_{A^0}V_{A^0} V_{\phi^0}>$ either \cite{Hatefi:2016enc}. Notice also to \cite{Hatefi:2012ve}, \cite{Park:2008sg}.} Hence, we make use of the Wick-like rule, to actually address all fermionic correlators.\footnote{
$x_{ij}=x_i-x_j$, and $\alpha'=2$.} One could read off doubling tricks, propagators  and all kinematical constraints from \cite{Hatefi:2016enc}. 

 \vskip.1in
To shorten the ultimate result of our scattering amplitude, we just illustrate its closed form  as follows
\beqa
{\cal A}^{C^{-2}\phi^0  \phi^0 T^0}&\sim& 2\Tr(\lam_1\lam_2\lam_3)\int dx_{1}dx_{2}
dx_{3}dx_{4}dx_{5}(P_{-}\fsC_{(n-1)}M_p)^{\al\be}I \xi_{1i}\xi_{2j}x_{45}^{-3/4}\nonumber\\&&\times
\bigg((i\alpha'k_{3c}) a^c_1 \bigg[ a^i_1a^j_2 -\eta^{ij} x_{12}^{-2}\bigg]-\alpha'^2  k_{2b}k_{3c}a^i_1 a^{cjb}_{2}\nonumber\\&&-\alpha'^2  k_{1a}k_{3c}a^j_2
 a^{cia}_{3}-i{\alpha'}^3 k_{1a}k_{2b}k_{3c}
a_4^{cjbia}\bigg)\labell{ampe1},\eeqa
where
\beqa
I&=&|x_{12}|^{\alpha'^2 k_1.k_2}|x_{13}|^{\alpha'^2 k_1.k_3}|x_{14}x_{15}|^{\frac{\alpha'^2}{2} k_1.p}|x_{23}|^{\alpha'^2 k_2.k_3}|
x_{24}x_{25}|^{\frac{\alpha'^2}{2} k_2.p}
|x_{34}x_{35}|^{\frac{\alpha'^2}{2} k_3.p}|x_{45}|^{\frac{\alpha'^2}{4}p.D.p},\nonumber\\
a^i_1&=&ip^{i}\bigg(\frac{x_{54}}{x_{14}x_{15}}\bigg),\nonumber\\
a^j_2&=&ip^{j}\bigg(\frac{x_{54}}{x_{24}x_{25}}\bigg),\nonumber\\
a^{c}_{1}&=&2^{-1/2}x_{45}^{-3/4}(x_{34}x_{35})^{-1/2} (\gamma^{c}C^{-1})_{\alpha\beta} ,\nonumber\\
a^{cjb}_{2}&=&2^{-3/2}x_{45}^{1/4}(x_{34}x_{35})^{-1/2}(x_{24}x_{25})^{-1}\bigg\{(\Gamma^{cjb}C^{-1})_{\alpha\beta}+\alpha' \eta^{bc} (\gamma^{j}C^{-1})_{\alpha\beta}\frac{Re[x_{24}x_{35}]}{x_{23}x_{45}}\bigg\}
,\nonumber\\
a^{cia}_{3}&=&2^{-3/2}x_{45}^{1/4}(x_{34}x_{35})^{-1/2}(x_{14}x_{15})^{-1}\bigg\{(\Gamma^{cia}C^{-1})_{\alpha\beta}+\alpha' \eta^{ac} (\gamma^{i}C^{-1})_{\alpha\beta}\frac{Re[x_{14}x_{35}]}{x_{13}x_{45}}\bigg\}.
\nonumber\eeqa
One finds the compact form of the last correlation function with different interactions as
\beqa
a_4^{cjbia}&=&
\bigg\{(\Gamma^{cjbia}C^{-1})_{{\alpha\beta}}+\alpha' h_1\frac{Re[x_{14}x_{25}]}{x_{12}x_{45}}+
\alpha' \eta^{ac}(\Gamma^{jbi}C^{-1})_{\alpha\beta}\frac{Re[x_{14}x_{35}]}{x_{13}x_{45}}+\alpha' \eta^{bc}(\Gamma^{jia}C^{-1})_{\alpha\beta}\nonumber\\&&\times\frac{Re[x_{24}x_{35}]}{x_{23}x_{45}}+\alpha'^2\eta^{ac}\eta^{ij}(\gamma^{b}C^{-1})_{\alpha\beta}\bigg(\frac{Re[x_{14}x_{35}]}{x_{13}x_{45}}\bigg)\bigg(\frac{Re[x_{14}x_{25}]}{x_{12}x_{45}}\bigg)
-\alpha'^2  \eta^{ab}\eta^{ij}(\gamma^{c}C^{-1})_{\alpha\beta}\nonumber\\&&\times
\bigg(\frac{Re[x_{14}x_{25}]}{x_{12}x_{45}}\bigg)^{2}
-\alpha'^2 \eta^{bc}\eta^{ij}(\gamma^{a}C^{-1})_{\alpha\beta} \bigg(\frac{Re[x_{14}x_{25}]}{x_{12}x_{45}}\bigg)\bigg(\frac{Re[x_{24}x_{35}]}{x_{23}x_{45}}\bigg)
\bigg\}\nonumber\\&&\times
2^{-5/2}x_{45}^{5/4}(x_{14}x_{15}x_{24}x_{25})^{-1}(x_{34}x_{35})^{-1/2}\label{hh22},\\
h_1&=&\bigg(\eta^{ab}(\Gamma^{cji}C^{-1})_{\alpha\beta}
+\eta^{ij}(\Gamma^{cba}C^{-1})_{\alpha\beta}\bigg).\nonumber
\eeqa

Now one is able to explicitly check the SL(2,R) invariance of the scattering amplitude, removing the volume of conformal killing group by fixing three positions of vertex operators as $ x_{1}=0 ,x_{2}=1,x_{3}\rightarrow \infty$. Finally we need to integrate out 2D complex integrals on upper half plane over the location of RR  (\cite{Fotopoulos:2001pt,Hatefi:2012wj,Hatefi:2016wof}).\footnote{
with \beqa
s&=&\frac{-\alpha'}{2}(k_1+k_3)^2, t=\frac{-\alpha'}{2}(k_1+k_2)^2, u=\frac{-\alpha'}{2}(k_2+k_3)^2, s'=s+\frac{1}{4},u'=u+\frac{1}{4}.
\nonumber\eeqa} Carrying all integrals out, using momentum conservation along the world volume of brane $(k_1+k_2+k_3)^a=-p^a$ and simplifying the scattering amplitude, one finds the final result  as
\beqa {\cal A}^{C^{-2} \phi^{0}  \phi^{0}T^{0} }&=&{\cal A}_{1}+{\cal A}_{2}+{\cal A}_{3}+{\cal A}_{4}
+{\cal A}_{5},\labell{fff}\eeqa
where
\beqa
{\cal A}_{1}&\!\!\!\sim\!\!\!&  2^{1/2}i (P_{-}\fsC_{(n-1)}M_p)^{\alpha\beta} 
\bigg[-\xi_1.\xi_2s'u' (k_1+k_2+k_3)_b (\gamma^{b}C^{-1})_{\alpha\beta}\bigg] L_1,
\nonumber\\
{\cal A}_{2}&\sim& -2^{1/2}2i (P_{-}\fsC_{(n-1)}M_p)^{\alpha\beta} 
\bigg[-k_{3c} (\gamma^{c}C^{-1})_{\alpha\beta} p.\xi_1 p.\xi_2+p.\xi_1 k_{2b}k_{3c} \xi_{2j}(\Gamma^{cjb}C^{-1})_{\alpha\beta}\nonumber\\&&+p.\xi_2 k_{1a}k_{3c} \xi_{1i}(\Gamma^{cia}C^{-1})_{\alpha\beta}-k_{1a}  k_{2b}k_{3c} \xi_{1i}
\xi_{2j}(\Gamma^{cjbia}C^{-1})_{\alpha\beta}\bigg](t+s+u+\frac{1}{2}) L_1,
\nonumber\\
{\cal A}_{3}&\sim&2^{1/2}2i (P_{-}\fsC_{(n-1)}M_p)^{\alpha\beta} 
\bigg[t p.\xi_1\xi_{2j}(\gamma^{j}C^{-1})_{\alpha\beta}+t k_{1a}\xi_{1i}\xi_{2j}(\Gamma^{jia}C^{-1})_{\alpha\beta}\bigg] L_2, \nonumber\\
{\cal A}_{4}&\sim&2^{1/2}2i (P_{-}\fsC_{(n-1)}M_p)^{\alpha\beta} 
\bigg[-t p.\xi_2\xi_{1i}(\gamma^{i}C^{-1})_{\alpha\beta}+t k_{2b}\xi_{1i}\xi_{2j}(\Gamma^{jbi}C^{-1})_{\alpha\beta}\bigg] L_2, \nonumber\\
{\cal A}_{5}&\sim&2^{3/2}i (P_{-}\fsC_{(n-1)}M_p)^{\alpha\beta} k_{3c}\bigg[-t \xi_{1i}\xi_{2j}(\Gamma^{cji}C^{-1})_{\alpha\beta}+2 k_{1a}k_{2b}\xi_1.\xi_2 (\Gamma^{cba}C^{-1})_{\alpha\beta}\bigg] L_2.\labell{cc1}\eeqa
where the functions
 $L_1,L_2$ are :
\beqa
L_1&=&(2)^{-2(t+s+u)}\pi{\frac{\Gamma(-u+\frac{1}{4})
\Gamma(-s+\frac{1}{4})\Gamma(-t+\frac{1}{2})\Gamma(-t-s-u-\frac{1}{2})}
{\Gamma(-u-t+\frac{3}{4})\Gamma(-t-s+\frac{3}{4})\Gamma(-s-u+\frac{1}{2})}},\nonumber\\
L_2&=&(2)^{-2(t+s+u)-1}\pi{\frac{\Gamma(-u+\frac{3}{4})
\Gamma(-s+\frac{3}{4})\Gamma(-t)\Gamma(-t-s-u)}
{\Gamma(-u-t+\frac{3}{4})\Gamma(-t-s+\frac{3}{4})\Gamma(-s-u+\frac{1}{2})}}.
\nonumber\eeqa

This amplitude in terms of field strength of RR, that is, $<V_{C^{-1}} V_{\phi^{-1}} V_{\phi^0}V_{T^0}>$  has not been computed at all. Making use of CFT we perform all the fermionic correlation functions as well as take into account the previous section. In order to shorten the computations, here we just determine the final result for the scattering amplitude as follows
\beqa {\cal A'}^{C^{-1}\phi^{-1}\phi^0 T^{0}}&=&{\cal A'}_{1}+{\cal A'}_{2}+{\cal A'}_{3},\labell{dd}\eeqa
with the following sub-amplitudes
\beqa
{\cal A'}_{1}&\!\!\!\sim\!\!\!& L_1\bigg[2\xi_{1i}\xi_{2j}k_{3c}k_{2b}
\Tr(P_{-}\fsH_{(n)}M_p\Gamma^{cjbi})-2\xi_{1i} p.\xi_2 k_{3c}\Tr(P_{-}\fsH_{(n)}M_p \Gamma^{ci})\bigg](-t-s-u-\frac{1}{2})
\nonumber\\&&+u's'\xi_1.\xi_2\Tr(P_{-}\fsH_{(n)}M_p) L_1,
\nonumber\\
{\cal A'}_{2}&\sim&-4 L_2  k_{3c}k_{2b} \xi_1.\xi_2 \Tr(P_{-}\fsH_{(n)}M_p \Gamma^{cb}),
\nonumber\\
{\cal A'}_{3}&\sim&2t L_2\Tr(P_{-}\fsH_{(n)}M_p\Gamma^{ji}) \xi_{1i}\xi_{2j}.
\label{ddd}\eeqa
$<V_{C^{-1}} V_{\phi^0} V_{\phi^0}V_{T^{-1}}>$  was already computed, we re-write it down as below   

\beqa {\cal A''}^{C^{-1}\phi^0\phi^0 T^{-1}}&=&{\cal A''}_{1}+{\cal A''}_{2}+{\cal A''}_{3},\labell{44}\eeqa
where
\beqa
{\cal A''}_{1}&\!\!\!\sim\!\!\!&iL_1\bigg[-2\xi_{1i}\xi_{2j}k_{1a}k_{2b}
\Tr(P_{-}\fsH_{(n)}M_p\Gamma^{jbia})-2 p.\xi_1 p.\xi_2\Tr(P_{-}\fsH_{(n)}M_p )\nonumber\\&&+2 p.\xi_{1}\xi_{2j}k_{2b}\Tr(P_{-}\fsH_{(n)}M_p \Gamma^{jb})+2 p.\xi_2 \xi_{1i} k_{1a}\Tr(P_{-}\fsH_{(n)}M_p \Gamma^{ia})\bigg](-t-s-u-\frac{1}{2})\nonumber\\&&
+iu's'\xi_1.\xi_2\Tr(P_{-}\fsH_{(n)}M_p)L_1,
\nonumber\\
{\cal A''}_{2}&\sim&-4i L_2 k_{1a}k_{2b}\xi_1.\xi_2\Tr(P_{-}\fsH_{(n)}M_p \Gamma^{ba}),
\nonumber\\
{\cal A''}_{3}&\sim&2it L_2 \Tr(P_{-}\fsH_{(n)}M_p\Gamma^{ji})\xi_{1i}\xi_{2j}.
\label{444}\eeqa
\vskip.1in

On the other hand, by applying momentum conservation along the world volume of brane we derive $s+t+u=-p^a p_a-\frac{1}{4}$.
Having considered the momentum conservation of an RR and a tachyon $k_a+p_a=0$, we would know  that  the quantity $p^ap_a$  should be sent to $k^2=-m^2$ mass of tachyon, that is,  $p^ap_a \rightarrow  \frac{1}{4}$. There is also non-zero coupling between two scalars and a gauge field  thus we come to a unique expansion as follows  $t\rightarrow 0, s\rightarrow -\frac{1}{4}, u\rightarrow -\frac{1}{4}$.
We also normalize the amplitude by  $(i\mu'_p\beta'\pi^{1/2})$ factor and consider the compact form of the expansions of functions as well.\footnote{
\beqa
L_1&=&-{\pi^{5/2}}\bigg( \sum_{n=0}^{\infty}c_n(s'+t+u')^n+\frac{\sum_{n,m=0}^{\infty}c_{n,m}[(s')^n(u')^m +(s')^m(u')^n]}{(t+s'+u')}\nonumber\\&&
+\sum_{p,n,m=0}^{\infty}f_{p,n,m}(s'+t+u')^p[(s'+u')^n(s'u')^{m}]\bigg),
\nonumber\\
L_2&=&-\pi^{3/2}\bigg(\frac{1}{t}\sum_{n=-1}^{\infty}b_n(u'+s')^{n+1}+\sum_{p,n,m=0}^{\infty}e_{p,n,m}t^{p}(s'u')^{n}(s'+u')^m\bigg).\labell{high67}\eeqa
Some of the coefficients are derived to be \beqa
&&b_{-1}=1,b_0=0,b_1=\frac{1}{6}\pi^2,b_2=2\z(3),e_{0,0,1}=c_1=\frac{\pi^2}{3},c_2=4\z(3),c_0=c_{1,0}=c_{0,1}=0.\nonumber
\eeqa}

\vskip.1in

For the simplicity let us first compare the scattering amplitudes  \reef{ddd} and \reef{444}.
Clearly all infinite $(t+s'+u')$- channel singularities are precisely re-generated which is why the last terms in   
${\cal A'}_{1}$ and ${\cal A''}_{1}$ are exactly the same. Using momentum conservation along the world volume of brane and replacing $k_{3c}=-(k_1+k_2+p)_c$ into ${\cal A'}_{2}$ as well as applying the Bianchi identity 
\beqa 
p_c \epsilon^{a_0...a_{p-2}bc}=0.\eeqa
 to ${\cal A'}_{2}$, we are able to exactly reconstruct all infinite t-channel singularities. Hence ${\cal A'}_{2}$ and ${\cal A''}_{2}$ are also equivalent. Now we turn to comparisons of all order contact terms. Indeed ${\cal A'}_{3}$ and ${\cal A''}_{3}$  are the same. The above argument and the same Bianchi identity holds for the first term of ${\cal A'}_{1}$, thus it is reproduced by the first term of ${\cal A''}_{1}$ as well. 
 
 \vskip.1in
 
We also use momentum conservation and replace $k_{3c}=-(k_1+k_2+p)_c$ into the second term of ${\cal A'}_{1}$ and finally apply the following Bianchi identity for RR's field strength to it
 \beqa
 \eps^{a_0\cdots a_{p}} \bigg( -p_{a_{p}} (p+1) H^{ij}_{a_0\cdots a_{p-1}}-p^j H^{i}_{a_0\cdots a_{p}}+p^i H^{j}_{a_0\cdots a_{p}}\bigg)&=&d H^{p+2}=0.
 \label{BI12}\eeqa
 or
 \beqa
   p_{a_{0}} \eps^{a_0\cdots a_{p}} \bigg( -p_{a_{p}} p (p+1) C_{ija_1\cdots a_{p-1}}-p^j
   C_{ia_1\cdots a_{p}}+p^i C_{ja_1\cdots a_{p}}\bigg)&=&0.\label{mm11}\eeqa

By doing so, one is able to reconstruct the 3rd and 4th terms of ${\cal A''}_{1}$. However, as it can be seen there is no chance to even produce the second term of ${\cal A''}_{1}$ from the result of $<V_{C^{-1}} V_{\phi^{-1}} V_{\phi^0}V_{T^{0}}>$. Thus we come to the fact that to be able to produce precisely all order $\alpha'$ contact terms of S-matrix elements of two scalar fields in the presence of RR, one needs to consider both scalar fields in zero picture. This fact is now confirmed by direct analysis of the above comparisons. Therefore $<V_{C^{-1}} V_{\phi^{0}} V_{\phi^0}V_{T^{-1}}>$ provides  very accurately all infinite contact terms for the scattering amplitudes.

In the next section we address our primary goals which  have to do with new contact interactions, singularity structures as well as restricted Bianchi identities. Finally we explicitly produce new string contact interactions  by  writing various new EFT couplings.

\section{All order contact interaction comparisons between $<V_{C^{-2}} V_{\phi^0} V_{\phi^0}V_{T^{0}}>$ and $<V_{C^{-1}} V_{\phi^0} V_{\phi^0}V_{T^{-1}}>$ }

Having applied momentum conservation to 4th term  ${\cal A}_{2}$ of \reef{cc1}, we would produce the 1st term 
${\cal A''}_{1}$. On the other hand, the 1st term ${\cal A}_{2}$ of \reef{cc1} can be written down as
\beqa
-2^{3/2}i \Tr(P_{-}\fsC_{(n-1)}M_p \gamma^{c}) (k_{1c}+k_{2c}+p_{c}) p.\xi_1 p.\xi_2  (t+s'+u')
L_1, \label{mm1}\eeqa
Using $p_c \fsC_{(n-1)}=\fsH_{(n)} $, one reveals that the 2nd term ${\cal A''}_{1}$ is exactly reconstructed. However,  the other two terms of \reef{mm1} are extra contact interactions and later on we consider their contributions towards approaching new restricted Bianchi identities.
We might rewrite the 2nd term ${\cal A}_{2}$ of \reef{cc1}  as follows

\beqa
2^{3/2}i \Tr(P_{-}\fsC_{(n-1)}\Gamma^{cjb} )
p.\xi_1 k_{2b}(k_{1c}+k_{2c}+p_{c}) \xi_{2j} (t+s'+u') L_1.\label{oo1}\eeqa
\reef{oo1}  is symmetric under exchanging $k_2$ and due to antisymmetric property of $\epsilon $ tensor, we come to know that 2nd term in \reef{oo1} does not have any contribution. The 3rd term in  \reef{oo1}  $(p\fsC=\fsH)$ generates exactly the 3rd term of ${\cal A''}_{1}$. Evidently the 1st term in \reef{oo1} will be extra contact interaction that will be taken into account later on.
Likewise, the same discussions hold for the 3rd term ${\cal A}_{2}$, so that it can be taken as 

\beqa
2^{3/2}i \Tr(P_{-}\fsC_{(n-1)}\Gamma^{cia} )
p.\xi_2 k_{1a}(k_{1c}+k_{2c}+p_{c}) \xi_{1i} (t+s'+u') L_1,\label{qq1}\eeqa
The 3rd term in above produces the 4th term ${\cal A''}_{1}$ and the 2nd term in \reef{qq1} will be  an extra contact term. Finally it can be shown that the 1st term ${\cal A}_{5}$ produces entirely ${\cal A''}_{3}$  as well as two extra contact terms as follows 

\beqa
2^{3/2}i t L_2\Tr(P_{-}\fsC_{(n-1)}M_p \Gamma^{cji}) (k_{1c}+k_{2c}+p_c) \xi_{1i}\xi_{2j}.\label{rr1}\eeqa

Therefore we were able to reconstruct all infinite contact interactions of  ${\cal A''}$. In addition to the contributions of the whole ${\cal A}_{3},{\cal A}_{4}$ of \reef{cc1}, we have also got various extra contact interactions from $<V_{C^{-2}} V_{\phi^{0}} V_{\phi^0}V_{T^{0}}>$.  One could write down all extra contact terms that are just present in $<V_{C^{-2}} V_{\phi^{0}} V_{\phi^0}V_{T^{0}}>$ as follows

\beqa
&&2^{1/2}2i (P_{-}\fsC_{(n-1)}M_p)^{\alpha\beta} 
\bigg[t p.\xi_1\xi_{2j}(\gamma^{j}C^{-1})_{\alpha\beta}+t k_{1a}\xi_{1i}\xi_{2j}(\Gamma^{jia}C^{-1})_{\alpha\beta}\nonumber\\&&-t p.\xi_2\xi_{1i}(\gamma^{i}C^{-1})_{\alpha\beta}+t k_{2b}\xi_{1i}\xi_{2j}(\Gamma^{jbi}C^{-1})_{\alpha\beta} \bigg] L_2, \label{uu1}\eeqa
as well as the following terms
\beqa
&&2^{3/2}i (t+s'+u') L_1\bigg[-\Tr(P_{-}\fsC_{(n-1)}M_p \gamma^{c}) (k_{1c}+k_{2c}) p.\xi_1 p.\xi_2  
 \nonumber\\&&+\Tr(P_{-}\fsC_{(n-1)}\Gamma^{cjb} )
p.\xi_1 k_{2b}(k_{1c}) \xi_{2j} 
+\Tr(P_{-}\fsC_{(n-1)}\Gamma^{cia} ) 
p.\xi_2 k_{1a}(k_{2c}) \xi_{1i} \bigg],\label{ss1}\eeqa

\beqa
2^{3/2}i t L_2\Tr(P_{-}\fsC_{(n-1)}M_p \Gamma^{cji}) (k_{1c}+k_{2c}) \xi_{1i}\xi_{2j}.\label{rr1}\eeqa
Now using some algebra, one realizes that the sum of  2nd and 4th  terms of \reef{uu1} cancels off precisely the entire contributions of \reef{rr1}.  Thus, we would have to just deal with the other above six extra contact interactions. Let us extract the traces, make use of antisymmetric property of Gamma Matrices and simplify further all the above extra contact terms that are just present in $<V_{C^{-2}} V_{\phi^{0}} V_{\phi^0}V_{T^{0}}>$
as follows

\beqa
 2^{3/2}i \xi_{1i} \xi_{2j} (t+s'+u') \frac{16 L_1}{(p-1)!}\bigg[-\frac{1}{p}(k_{1c}+k_{2c}) p^i p^j  \epsilon^{a_{0}...a_{p-1}c} C_{ a_{0}...a_{p-1}}\nonumber\\
  +k_{2b} k_{1c} \epsilon^{a_{0}...a_{p-2}bc}\bigg(p^i C^{j}_{ a_{0}...a_{p-2}}- p^j C^{i}_{ a_{0}...a_{p-2}} \bigg)\bigg],\label{669}\eeqa

\beqa
&&2^{3/2}i t\xi_{1i} \xi_{2j}\frac{16 L_2}{(p+1)!} \epsilon^{a_{0}...a_{p}}\bigg(p^i C^{j}_{ a_{0}...a_{p}}- p^j C^{i}_{ a_{0}...a_{p}} \bigg). \label{778}\eeqa

The appearance of the 1st term in \reef{669} is essential, because it is symmetric under interchanging scalar fields and it can be reconstructed by Taylor expansion of scalar fields. 

\vskip.2in

In this section,  we show that one needs to explore new EFT couplings  to be able to produce all the extra contact terms of scattering amplitude (\reef{669} and \reef{778} terms) that are just appeared in 
$<V_{C^{-2}} V_{\phi^{0}} V_{\phi^0}V_{T^{0}}>$.

\vskip.2in

In order to produce the 1st term in 2nd line of \reef{669} one needs to write down a new  EFT coupling as follows.

 Suppose the 1st scalar field comes from Taylor expansion and the 2nd scalar and tachyon comes from 
 sort of new structure (neither pull-back nor Myers term) so that they cover the entire world volume spaces as follows
 
 \beqa
i(2\pi\alpha')^3\beta'\mu'_p \int d^{p+1}\sigma 
\Tr(\partial_i C^j_{p-1}\wedge D\phi_j \wedge DT\phi^i),
\labell{pp1}
\eeqa

Extracting the above coupling we seem to arrive at
\beqa
\frac{i(2\pi\alpha')^3\beta'\mu'_p}{(p-1)!} \int_{\Sigma_{p+1}} d^{p+1}\sigma 
\Tr(\partial_i C^j_{a_0...a_{p-2}} D_{a_{p-1}}\phi_j D_{a_{p}}T\phi^i)   \epsilon^{a_{0}...a_{p}}.
\labell{pp1b}
\eeqa
If we take the above coupling to momentum space we obtain
\beqa
\xi_{1i}\xi_{2j} p_i C^j_{a_0...a_{p-2}} k_{1b} k_{3c} \epsilon^{a_0...a_{p-2}bc}.\eeqa

Using momentum conservation $k_{3c}=-(k_{1c}+k_{2c}+p_c)$  and restricted Bianchi identity $p_c\epsilon^{a_0...a_{p-2}bc}=0$, we were able to derive the 1st contact term of 2nd line of \reef{669} and the 2nd term could also be produced by the same argument as well as exchanging the transverse indices. Note that by applying properly higher derivative corrections one can promptly generate their all order $\alpha'$ corrections (see section 5,6 of \cite{Hatefi:2012wj}).

\vskip.1in

We have also some non-zero contact interactions for $n=p+3$ case as follows

\beqa
&&2^{3/2}i t\xi_{1i} \xi_{2j}\frac{16 L_2}{(p+1)!} \epsilon^{a_{0}...a_{p}}\bigg(p^i C^{j}_{ a_{0}...a_{p}}- p^j C^{i}_{ a_{0}...a_{p}} \bigg),\nonumber\eeqa

One might wonder whether or not the above couplings can be produced by Bianchi identity that has been proposed for potential RR form field in \reef{mm11}. Clearly in that identity full $(p+1)$ world volume dimensions have been covered up  so the only way to reconstruct  couplings appeared in \reef{778} is through imposing the following structure in an EFT coupling 

\beqa
\frac{i(2\pi\alpha')^3\beta'\mu'_p}{2(p+1)!} \int_{\Sigma_{p+1}} d^{p+1}\sigma 
\bigg[\partial_i C^j_{a_0...a_{p}} -\partial_j C^i_{a_0...a_{p}}\bigg](\phi^i\phi^jT) \epsilon^{a_{0}...a_{p}}.
\labell{bbcz}
\eeqa

The coefficient \reef{bbcz} is fixed at $\alpha'^3$ order to be able to match it up with its own string part. One needs  to highlight the following fact about scattering amplitude potential. Given the precise and direct form of the amplitude,  the coefficients of all above couplings are exact as they involve no ambiguities at all.  
Hence, among  new couplings   in this section we confirmed the claim of  known dielectric effect   \cite{Myers:1999ps} that almost all transverse scalar fields or background fields in supergravity analysis are functions of superYang-Mills.

\vskip.1in

If we insert the expansion of $L_1$ inside of ${\cal A}_{2}, {\cal A}_{1}$ of \reef{cc1} accordingly, one can show that  $\alpha'^3$ order of those contact interactions are obtained by employing either Taylor expansion, pull-back or mix combination of them as follows

\beqa
\frac{(2\pi\alpha')^3\beta'\mu'_p}{2p!}\int d^{p+1}\sigma\,
\veps^{a_0\cdots a_p}\bigg( \Tr(\partial_{a_0}T
\phi^i\phi^j)\prt_i\prt_jC_{a_1\cdots a_p}
+2p\Tr(\prt_{a_0}T\prt_{a_1}\phi^i\phi^j)\,\prt_j C_{ia_2\cdots a_p}\nonumber\\+(p-1)p\Tr(\partial_{a_0}T\prt_{a_1}
\phi^i\prt_{a_2}\phi^j)\,C_{ija_3\cdots a_p}\bigg),
\labell{s22}
\eeqa

The following new EFT coupling is also needed
\beqa
\frac{i(2\pi\alpha')^3\beta'\mu'_p}{4p!}\int d^{p+1}\sigma \veps^{a_0\cdots a_p}
\Tr(\partial_{a_0}T\phi^i\phi_i)\,C_{a_1\cdots a_p}.
\labell{snjh22}
\eeqa
which does not come from any standard effective field theory at all. 

If one substitutes the expansion of $L_2$ inside the 1st term of  ${\cal A}_{5}$, then  one realizes that not only $\alpha'^3$ order but also the mix combination of  commutator of scalar fields and covariant derivative of tachyon is also needed as follows

\beqa
\frac{(2\pi\alpha')^3\beta'\mu'_p}{4p!}\int d^{p+1}\sigma \veps^{a_0\cdots a_p} \Tr(D_{a_0}T[\phi^i,\phi^j]) C_{jia_1\cdots a_p},
\labell{s89h267}
\eeqa

We can extract the above coupling out and rewrite it  as follows
\beqa
\frac{(2\pi\alpha')^3 \beta'\mu'_p}{2p!}\int d^{p+1}\sigma \veps^{a_0\cdots a_p}
 \Tr(\partial_{a_0}T
\phi^i\phi^j) C_{jia_1\cdots a_p}.
\nonumber
\eeqa

Once more this term does not come from any standard effective field theory either. Now one is able to properly switch on all higher derivative corrections  on above new EFT couplings of type IIA,IIB and find out their all order $\alpha'$ corrections.

\section{Singularity comparisons }

Unlike the amplitude of $<V_{C^{-2}} V_{\phi^0} V_{\phi^0}V_{A^{0}}>$ that had 
an infinite u,s -channel BPS bulk singularities, for this particular $<V_{C^{-2}} V_{\phi^0} V_{\phi^0}V_{T^{0}}>$ of non-BPS branes we come to know that neither there are u,s- channel bulk singularities nor any other u,s -channel singularities. Indeed  based on the selection rules for non-BPS  branes (which are special kinds of  Chan-Paton factor like matrices \cite{Hatefi:2013yxa}), there is no coupling between two scalars and a tachyon. This is a promising reason towards not having any s,u channel poles  for this  $<V_{C^{-2}} V_{\phi^0} V_{\phi^0}V_{T^{0}}>$.

\vskip.1in

To start comparing singularity structures, we try to first re-generate all singularity structures that have been shown up in \reef{444}. If we use momentum conservation and $p_c \fsC_{(n-1)}=\fsH_{(n)}$
to the entire ${\cal A}_{1}$ of \reef{fff}, then we are able to exactly reconstruct all order $(t+s'+u')$ channel singularities of  \reef{444}.  In the other words,  the 5th term of ${\cal A''}_{1}$ will be re-produced.
On the other hand, by applying momentum conservation, using the Bianchi identity $p_c \epsilon^{a_0...a_{p-2}bc}=0$ and taking $p_c \fsC_{(n-1)}=\fsH_{(n)}$ to the 2nd term of ${\cal A}_{5}$ of \reef{fff}, one is immediately able to generate all infinite t-channel singularity structures of ${\cal A''}_{2}$ as well.

\vskip.1in

Let us briefly produce all infinite t-channel gauge field singularities as well as some other new EFT couplings. The second term in ${\cal A}_{5}$ dictates us the following singularity structures:
\beqa
{\cal A}_{t-channel}&=&\frac{64i\mu'_p\beta'\pi^{2}}{t(p-2)!} p_c C_{a_{0}\cdots a_{p-3}} \xi_1.\xi_2k_{1a}k_{2b}
\eps^{a_{0}\cdots a_{p-3}cba}\sum_{n=-1}^{\infty}b_n(u'+s')^{n+1}.  
\labell{masspole}\eeqa

The sub-amplitude in field theory is 
 \beqa
{\cal A}&=&V_\alpha^{a}(C_{p-2},T_3,A)G_{\alpha\beta}^{ab}(A)V_\beta^{b}(A,\phi_1,\phi_2),\labell{amp54}\eeqa

Gauge field propagator is found from its kinetic term $ (2\pi\alpha')^2  \Tr(F_{ab} F^{ab})$ that has been fixed in DBI action, and 
$V_\beta^{b}(A,\phi_1,\phi_2)$ is read off from the kinetic term of scalar fields  $\frac{ (2\pi\alpha')^2}{2}  \Tr(D_a\phi_i D^a\phi^i)$, that is also fixed because there are no corrections to kinetic terms. $V_\alpha^{a}(C_{p-2},T_3,A)$ vertex  operator can be extracted from Wess-Zumino couplings as follows
\beqa
2\mu'_p\beta'(2\pi\alpha')^{2}\Tr(C_{a_0...a_{p-3}} F_{a_{p-2}a_{p-1}} D_{a_p}T) \eps^{a_{0}\cdots a_{p}},\label{ee414}\eeqa
To derive all infinite t- channel poles, one needs to reconstruct all order $\alpha' $ higher derivative corrections to \reef{ee414}. Based on the facts that kinetic terms and propagators do not receive any corrections, one applies all order $\alpha'$ corrections to  \reef{ee414}  as follows 
\beqa
2\mu'_p\beta'(2\pi\alpha')^{2} \sum_{n=-1}^{\infty}b_n\Tr(C_{a_0...a_{p-3}} \wedge D_{a_0}... D_{a_n}F \wedge D^{a_0}... D^{a_n}DT).\label{ee44}\eeqa
 Keeping in mind that the following vertices are extracted from an EFT

\beqa
V_\alpha^{a}(C_{p-2},T_3,A)&=&2\mu'_p\beta'(2\pi\alpha')^{2}\frac{\Tr(\lam_3\Lambda^{\alpha})}{(p-2)!} \eps^{a_{0}\cdots a_{p-2}ac} p_c C_{a_{0}\cdots a_{p-3}}k_{a_{p-2}}\sum_{n=-1}^{\infty}b_n(s'+u')^{n+1},\nonumber\\
V_\beta^{b}(A,\phi_1,\phi_2)&=& iT_p(2\pi\alpha')^{2} \xi_{1}.\xi_{2}(k_1-k_2)^{b} \Tr(\lambda_{1}\lambda_{2}\Lambda_{\beta}),
\label{mmnn1}\\
G_{\alpha\beta}^{ij}(A)&=&\frac{i\delta_{\alpha\beta}\delta^{ab}}{(2\pi\alpha')^{2}T_p(t)}.\nonumber\eeqa

with $k$ as an off-shell momentum of gauge field and replacing them in \reef{amp54}, one is precisely able to produce all infinite t-channel singularities of string amplitude  without any residual contact term. If we consider the second part of $L_1$ expansion inside its string counterparts, we clarify that  a new coupling is needed to actually reproduce the rest of contact terms. In fact, the structure of this coupling is made out of two covariant derivatives of scalar fields so that neither they come from pull-back,Taylor expansion  nor from commutator of scalar fields. Namely, string amplitude indicates that in order to cover the full $(p+1)$ dimensions of world volume spaces, scalar fields must enjoy the following structure  

\beqa \beta'\mu'_p (2\pi\alpha')^3 \int_{\Sigma_{p+1}} d^{p+1}\sigma C_{p-2}\wedge D\phi^i\wedge D\phi_i\wedge DT.\label{gh1}\eeqa
The coefficient of the above coupling is fixed so that the leading contact interaction can precisely reconstruct the leading order of string parts at 3rd order of $\alpha'$. Given the entire explanations on all order $\alpha'$ higher derivative corrections in \cite{Hatefi:2012rx}, one can promptly apply all the corrections to \reef{gh1} and start to fix without any ambiguities all order corrections as well.
 
On the other hand, all infinite tachyon singularities  ${\cal A}_{1}$  can be structured by  the following sub-amplitude in an EFT
\beqa V^{\alpha}(C_{p},T)G^{\alpha\beta}(T)V^{\beta}(T,T_3,\phi_1,\phi_2),\nonumber\eeqa

One needs to employ  the tachyon propagator, $2i\beta'\mu'_p (2\pi\alpha') \int_{\Sigma_{p+1}} d^{p+1}\sigma C_{p}\wedge DT$  coupling and all order corrections of two scalar two tachyon couplings derived in \cite{Hatefi:2012wj} to be eventually able to produce all the singularities as follows

\beqa
&&-16i\xi_1.\xi_2u's'\pi\alpha'^{2}\beta'\mu_p'\frac{ \eps^{a_{0}\cdots a_{p-1}b} p_b C_{a_{0}\cdots a_{p-1}}}{p!(t+s'+u')}
\sum_{n,m=0}^{\infty} (a_{n,m}+b_{n,m})(s'^{m}u'^{n}+s'^{n+1}u'^{m}).
 \labell{amphigh11}\eeqa

As a final comment, we would like to point out  the importance of new EFT couplings that are discovered in this paper. Not only they will have applications to pure D-brane areas  but also would have substantial effects in predicting mathematical results/identities or symmetries beyond string amplitudes. 
Given our long endeavours, we hope to be able to diversify and fully address various open questions in this field in near future.

\section{Conclusions}

 First of all the derivations of  the entire form of one potential C-field, two transverse scalar fields and a tachyon  $<V_{C^{-2}}  V_{\phi^{0}} V_{\phi ^{0}}  V_{T ^{0}}>$ and  $<V_{C^{-1}}  V_{\phi^{-1}} V_{\phi ^{0}}  V_{T ^{0}}>$ on both IIA,IIB  have been found out. We also showed that unlike  $<V_{C^{-2}}  V_{\phi^{0}} V_{\phi ^{0}}  V_{A ^{0}}>$ for this particular non-BPS amplitude, there are no  u,s channel bulk singularity structures either.  All infinite t, $(t+s'+u')$  singularities have been shown to match with an effective field theory. 
 
\vskip.1in
In section 3,  we showed that one needed  to explore new EFT couplings  to be able to produce all the extra contact terms of scattering amplitude (\reef{669} and \reef{778} terms) that are just appeared in 
$<V_{C^{-2}} V_{\phi^{0}} V_{\phi^0}V_{T^{0}}>$. These contact terms are the terms that carry momentum of RR in bulk or transverse directions and can not be derived from the other analysis. Making use of direct string amplitude calculations, we have also confirmed the presence of new EFT couplings as
 
 \beqa
 &&(2\pi\alpha')^3\mu'_p \beta' \int_{\Sigma_{p+1}} d^{p+1} \sigma  \Tr(\partial_iC^{j}_{p-1} \wedge D\phi_j\wedge D T\phi^i),  \nonumber\\&& (2\pi\alpha')^3\mu'_p \beta' \int_{\Sigma_{p+1}} d^{p+1} \sigma \Tr(C_{p}\wedge DT \phi^i\phi_i), \nonumber\\&&
(2\pi\alpha')^3\mu'_p \beta'  \int_{\Sigma_{p+1}} d^{p+1} \sigma\Tr(C_{p-2} \wedge D\phi^i\wedge D\phi_i\wedge DT),  \nonumber\\&&
  (2\pi\alpha')^3\mu'_p \beta' \int_{\Sigma_{p+1}} d^{p+1} \sigma\Tr(D_{a_0}T[\phi^i,\phi^j])C_{jia_1\cdots a_p}\veps^{a_0\cdots a_p}.  \nonumber
 \eeqa

  where none of these couplings comes from standard effective field theory methods nor from pull-back, Taylor expansion. Notice to the important point as follows. We clarified the structures of new couplings on the entire space-time dimensions, fixed their coefficients without any ambiguities. More crucially, the generalization of their all order $\alpha'$ higher derivative corrections without any ambiguity has been constructed as well. Ultimately, the presence of the commutator of  two transverse scalar fields from the exponential of Wess-Zumino action even for non-BPS branes has also been proven.

\section*{Acknowledgements}

Parts of the computations of this paper have been taken place during my visits at Oxford University (Oxford Holography and Mathematics Group ), Swansea University  and I would like to thank L. Mason, A. Starinets, R. Penrose, C. Nunez, P. Kumar and G. Shore for valuable discussions as well as very warm hospitality.  The author would like to thank P. Anastasopoulos, M.Douglas, C.Hull, R. Myers, M. Gary, J.Knapp,  T. Wrase,  A. Rebhan, H. Steinacker, C. Papageorgakis and  D. Young  for  useful discussions and comments. He would also like to stress that parts of the computations of this paper  were carried out during his 2nd  post doc at string theory group at Queen Mary University of London and it is a pleasure to thank QMUL  and its string group for supports and enjoyable discussions throughout his presence. This work was supported by the FWF project P26731-N27.

  \end{document}